# Savu: A Python-based, MPI Framework for Simultaneous Processing of Multiple, N-dimensional, Large Tomography Datasets

Nicola Wadeson, Mark Basham
Diamond Light Source Ltd.
Didcot, Oxfordshire, OX11 0DE, UK
nicola.wadeson@diamond.ac.uk
mark.basham@diamond.ac.uk

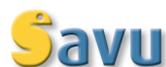

*Abstract*—Diamond Light Source (DLS), the UK's synchrotron facility, attracts scientists from across the world to perform ground-breaking x-ray experiments. With over 3000 scientific users per year, vast amounts of data are collected across the experimental beamlines, with the highest volume of data collected during tomographic imaging experiments. A growing interest in tomography as an imaging technique, has led to an expansion in the range of experiments performed, in addition to a growth in the size of the data per experiment.

Savu is a portable, flexible, scientific processing pipeline capable of processing multiple, n-dimensional datasets in serial on a PC, or in parallel across a cluster. Developed at DLS, and successfully deployed across the beamlines, it uses a modular plugin format to enable experiment-specific processing and utilises parallel HDF5 to remove RAM restrictions. The Savu design, described throughout this paper, focuses on easy integration of existing and new functionality, flexibility and ease of use for users and developers alike.

## I. INTRODUCTION

Diamond Light Source (DLS) is the UK's national synchrotron science facility. A synchrotron is a particle accelerator capable of accelerating electrons to close to the speed of light. Magnets cause these electrons to change direction and consequently to release energy in the form of x-rays, which are directed towards many experimental stations surrounding the accelerator, known as beamlines. X-rays produced in this way are of incredible brilliance, many orders of magnitude greater than laboratory or hospital sources, which is the main reason for the vast quantities of data collected. Once inside the beamline, the x-rays are conditioned for scientific use: The wavelength is selected, the beam is shaped and focused and, finally, directed towards the location of the sample.

Scientists from across the world come to DLS to conduct a range of experiments to perform cutting edge research. They use a variety of techniques, such as spectroscopy, macromolecular crystallography and imaging, to conduct research from a broad array of fields, from pharmaceuticals to materials science. The outcome of this research has led to ground-breaking discoveries, from new medicines and treatments for disease to innovative engineering and cutting-edge technology.

With over 3000 scientific users visiting DLS each year, vast amounts of raw data are output from the experiments (currently over 2 Petabytes a year and rising). The highest volume of data is collected during tomographic imaging experiments, where high-resolution images (or projections) are continuously captured as a small sample rotates, from 0 to 180 degrees (a single scan), on a stage positioned within the beam. The number of datasets (measurements) collected per scan and the number of dimensions per dataset depends on the type of data being collected.

The tomography community at DLS is growing in popularity, with five beamlines now applying the approach alongside the two initial imaging beamlines (I12 and I13) [1, 2]. The amount of data collected per tomography experiment is also growing, as improvements in data acquisition and computer technology lend the technique to time-resolved imaging. By capturing multiple tomography scans continuously, and subsequently adding an extra time dimension to the data, a whole new range of dynamic experiments can be performed, such as granular and fluid flow, or stress and strain experiments.

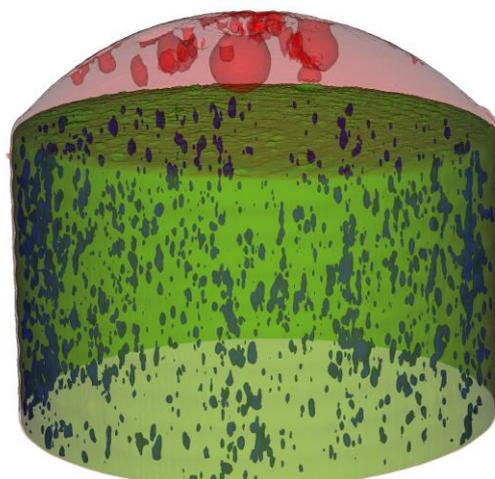

Fig.1: This rendering of data reconstructed using Savu shows a 500 micron diameter Aluminum (green) Magnesium (blue) alloy pin, with a droplet of sea water (red) on the top of the pin. The study was a time resolved experiment looking at the corrosion on the surface of the pin (too small to see in this rendering) and the corresponding Hydrogen bubbles in the sea water (visible at the top of the droplet).

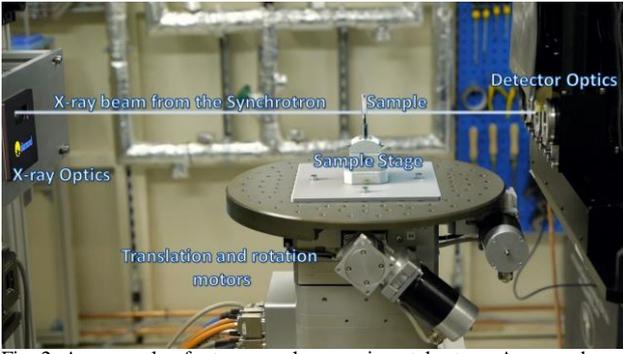

Fig. 2: An example of a tomography experimental setup. An x-ray beam traverses a sample, placed on a rotation stage, and interacts with a detector, which records the x-ray beam information.

The size of the raw data is experiment dependent, but as an indication, a typical full-field transmission tomography single scan has approximately 3k *x* 4k *x* 4k values $\approx$ 96GB, so in this experimental setup the raw data has size 96GB x N, where N is the number of scans. Tomography data from I12 and I13, which collect this type of data, accounts for around 60% of the data ever collected at DLS. The raw data is stored as 16 bit unsigned integer values, and the size is immediately doubled on processing, providing a challenge for the Data Analysis scientists who provide the software to the beamlines.

## II. TOMOGRAPHIC IMAGING AT DIAMOND

### A. Data Processing

The aim of any tomographic experiment is to produce at least one, 3-D computerized (tomographic) image of the sample (Fig. 1), where the values represent the distribution of some physical quantity (dependent upon the type of measurement(s) collected). This allows the investigator to see inside the sample non-destructively.

In theory, tomographic image reconstruction is relatively straightforward; requiring a simple correction, linearisation and standard filtered back-projection (FBP) reconstruction [3], and this is what was provided, as a GPU, cluster-based code, to beamline users at Diamond, to process **single** tomography scans. However, in reality, the experimental setup adds artefacts into the reconstructed image and the data requires some filtering to remove them. Experiments are becoming more complex, requiring processing of multiple, n-dimensional datasets, often simultaneously, and significant pre-processing before image reconstruction is possible [4].

This additional processing is often performed in the form of small scripts, written by beamline staff, which is inefficient, lacking cohesion, and requires them to have significant experience with the computation facilities that are available at DLS. In addition to this, beamlines users often have little or no knowledge of tomographic reconstruction, so the task of supplying them with software that is user friendly, flexible and gives better quality reconstructions, without overwhelming them, is not trivial. This fragmented approach to processing also has implications on processing speed. It is important that the speed of data processing is in line with the speed of data collection so, ideally, the results of previous data could help to drive the course of the experiments. Users should finish their experiments will all their data processed to the highest quality.

### B. Geometry and Data Collection

There are two types of data collection at DLS that can produce tomographic data: Full-field imaging and mapping scans. In full-field imaging, a sample is placed between the source of the x-ray beam and an x-ray detector, with the sample fully illuminated by the beam, as shown in Fig. 2. A parallel beam of x-rays traverses the object and the attenuated x-ray beam is recorded in the detector, producing a 2-D discrete sampling of the x-ray beam intensity, at equally spaced points on a 2-D plane, perpendicular to the beam.

This 2D array gives a 'projection' image of the object, known as projection data. The sample is rotated on a stage and the projection data is recorded at regular, angular intervals. The projection data is stacked in a third dimension to produce a single tomography scan raw dataset, with the final output as a 3-D HDF5 dataset [5] inside a NeXus file [6]. Fig. 3 shows the data processing steps in the 3-D coordinate system (x, y, θ). By slicing the data in the θ-dimension, the data can be processed as projections. Each projection requires pre-processing before the data can be sliced in the y-dimension to produce a stack of sinograms, which can be independently reconstructed producing 2-D slices of the reconstructed volume. These slices can be stitched back together to produce

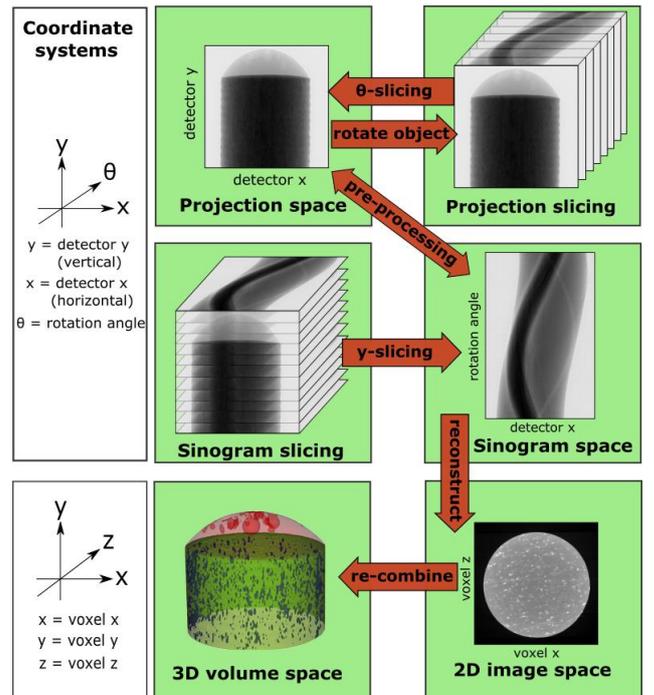

Fig.3: An illustration of the data spaces involved in the processing of full-field tomography data.

a 3D computerised image of the sample.

The parallel geometry of the synchrotron x-ray beam, and subsequent data collection, lends itself naturally to parallel processing. In reality, additional processing steps would be required before reconstructing and these could be applied in sinogram or projection space, depending on what the methodology required. Therefore, any processing software must allow easy and efficient transition from projection to sinogram processing, and careful considerations must be placed on how to organised the data in memory.

Mapping scans give rise to multi-modal data collection [7]. A sample is placed on a rotating stage in the beam, but the sample is rotated and translated through a thin pencil beam of x-rays to build up the same object coverage as in the full-field case [8]. With this setup it is possible to collect more information along the line of integration through the object, such as x-ray diffraction and fluorescence data as well as absorption data, Fig. 4. Not only is the data collection more complicated, but the absorption, fluorescence and diffraction datasets are 3-D, 4-D and 5-D respectively. Subsequently, additional data access patterns are required, such as 'spectra' and 'diffraction', to go along with 'projection' and 'sinogram', which are no longer necessarily 3-D. Additionally, it is useful to correct fluorescence data with the absorption data, requiring multiple datasets to be processed at the same time.

### C. Introduction to Savu

Given the evolving tomography requirements at Diamond, the previous tomography software had become outdated. With this in mind, a new in-house tomography reconstruction pipeline, called Savu, was theorised in detail in the paper [4], along with a comprehensive list of challenges faced at Diamond when processing tomography data. Since then, Savu has been implemented, officially released and rolled out across the beamlines, and successfully deals with reconstructing up to 3TB datasets, with no obvious barriers in moving to larger datasets once they are collected. The remainder of this paper aims to detail the design and implementation of this framework.

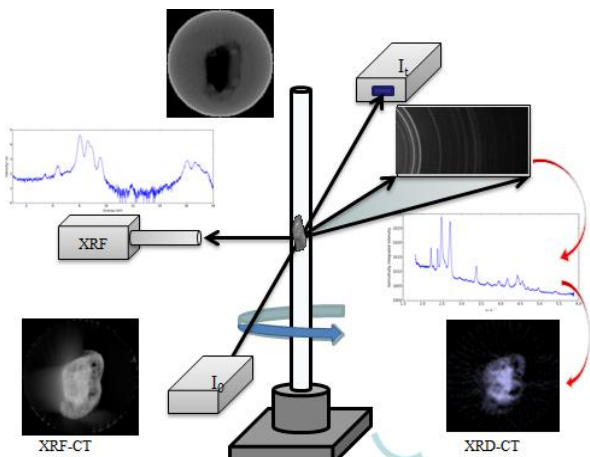

Fig. 4: An example of multi-modal experimental setup. The sample is rotated and translated through a thin pencil beam of x-rays to build up fluorescence, diffraction and absorption tomography datasets.

### III. SAVU DESIGN

Savu [9] is an open-source, tomography reconstruction and processing pipeline, developed at DLS. It is an object-oriented, python2-based code [10,11,12], capable of processing multi-modal, n-dimensional data in serial on a PC, or in parallel across a cluster, using the Message Passing Interface (MPI) [13]. The Savu framework design is focused on easy integration of existing and new functionality, flexibility and ease of use for users and developers alike. It uses a modular, 'plugin' format, where the movement and organisation of the data is hidden from the plugin developer. In this sense the core framework and the plugins are separate, so the plugin developer does not need to understand how the data is organised and cannot destabalise the core framework.

Each plugin performs an independent processing task and can be selected from a repository to form part of a process list, passed to Savu at runtime, enabling each run to be tailored to a specific experiment. By passing this process list, along with a data file and output folder to Savu at runtime, a processing chain is initiated. First, a plugin list check is performed on the data, highlighting any inconsistencies in the process list, followed by a setup phase and then the main plugin processing is performed. A full description of this processing chain is given throughout the following sections, alongside the series of framework images, Figs. 5-7,10 (key in Fig. 1), with the main Savu concepts required for this description given below.

### A. Handling Big Data

In order to handle multiple, n-dimensional, very large datasets (often simultaneously) the framework runs across multiple nodes of a cluster using MPI. It utilises parallel HDF5 [5], and subsequently parallel I/O, to read and write data directly to and from file at high speeds, which removes RAM-based memory restrictions. Therefore, each dataset, loaded or created during a Savu run, relates to a separate HDF5 file. When a plugin performs operations on an input dataset, the output result is stored in a separate file, and all of these files are linked together inside a Nexus file, allowing the result of intermediate processing steps to be viewed alongside the final result (e.g. using DAWN [14]). Savu allows these intermediate files to be stored in a separate location if required, so that they can be removed to reduce storage requirements, or to make use of fast but volatile temporary storage.

### B. Datasets

A dataset, inside the Savu framework, actually refers to a 'Data' object. Each Data object must contain the following

- A link to a data source
- Name
- Shape
- Axis labels
- Data access patterns

The internal data files are usually HDF5, but initial data may be different, such as a stack of tiff images. Each Data object also has a 'metadata' dictionary, where any extra information associated with the dataset can be populated, such as physical units and geometric information. There are two types of datasets within the framework: `in_datasets` and `out_datasets`. An `in_dataset` contains a data file that is available for processing, and an `out_dataset` contains a data file that is yet to be populated. Each `in_dataset` is uniquely identified by its name. If an `out_dataset` is created with the same name then, once it has been populated, it will replace the `in_dataset` to become the available dataset for the next step in the processing chain and the `in_dataset` will no longer be available for processing.

*C. Data Access Patterns*

Each dataset can have multiple data access 'patterns' associated with it. For instance, a 3-D tomography dataset has projection and sinogram patterns, which require the data to be sliced in the θ and y directions respectively (Fig. 3). A 'pattern' determines how the data should be sliced and is defined to have core and slice dimensions. For example, in the nx_tomo_loader.py plugin, the patterns associated with a 3-D dataset are:

```
tomo_3D.add_pattern('PROJECTION',
                    core_dir=(x, y),
                    slice_dir=(θ,)),

tomo_3D.add_pattern('SINOGRAM',
                    core_dir=(x, θ),
                    slices_dir=(y,)),
```

but with a 4-D dataset are:

```
tomo_4D.add_pattern('PROJECTION',
                    core_dir=(x, y),
                    slice_dir =(θ, scan)),

tomo_4D.add_pattern('SINOGRAM',
                    core_dir=(x, θ),
                    slice_dir =(y, scan)),
```

where x, y, θ and scan are mapped to their respective data dimensions (e.g.θ =0, y=1, x=2, scan=3 in this instance). A frame of data is then all elements in each core dimension and one in each slice dimension.

A plugin can specify any pattern, associated with a requested dataset, for processing as well as a number (*m*) of frames. The main plugin processing function inside the plugin will then receive *m* frames at a time until all the data has been processed. If there are multiple slice dimensions, the data will be sliced in the order specified in the slice_dir entry, so the first stated dimension will be the fastest changing dimension. This way, different loaders can associate the same pattern names with data, regardless of the data dimensionality or axis ordering, and the plugin will receive the same data, at each processing step, regardless. However, each pattern with the same name must have the same number of core dimensions, but different patterns can be created with differing numbers of core dimensions. For instance, if the data had dimensions (θ, y, x, E), where E is energy, then a spectrum pattern could be assigned to the dataset,

```
tomo_4D.add_pattern('SPECTRUM',
                    core_dir=(E),
                    slices_dir=(θ, y, x)),
```

*D. Core framework*

The core framework runs and controls the processing chain, and is responsible for the management of the datasets. It is capable of holding (and processing) multiple datasets at a time; creating and deleting them as the processing chain is traversed, with each dataset experiencing its own, unique list of processing steps. The core framework also deals with the movement and organisation of the data, including data slicing and data padding. It is responsible for the transfer of data frames, requested and returned by a plugin, to and from the relevant files.

*E. Processing chain (process list)*

A process list determines the chain of processing during a Savu run. It is created using a simple command line tool, called a configurator, and contains a list of plugins that should be applied to the data, along with access to plugin parameters that can be tailored for a specific experiment. The `in_datasets` and `out_datasets` parameters are set here, to determine which datasets a plugin should be applied to. Each dataset can experience its own, unique list of processing steps, as not all plugins need to be applied to all datasets and vice-versa. Each processing chain should start with at least one loader plugin and end with a saver.

*F. Plugins*

Plugins are individual entities in the form of a python module, which must contain one python class of the same name and inherit from a base class that determines its type (e.g. BaseFilter, BaseRecon, which will be collectively referred to as BaseType) and a plugin driver for the CPU or GPU. There are three different types of plugins: loaders; savers and processing plugins.

*1) Plugin drivers*

The plugin drivers control the calls to the plugin processing methods (see *Main Processing* below) and are important when running MPI Savu as they determine the number of MPI processes that will execute a plugin. The CPU driver allows all processes to process the plugin; however the GPU driver creates an MPI communicator for a reduced number of processes, equal to the number of available GPUs, and the remaining processes must wait until the plugin has completed to resume their processing.

*2) Loaders and Savers*

Loader and saver plugins are two special types of plugins that do not perform any data processing. A link to a data file or folder (usually raw data), is passed to the pipeline at runtime and it is the job of a 'loader' plugin to create the associated Data objects including assigning unique names. Multiple datasets can be created per loader, and a process list can contain multiple loaders.

The collection of datasets created by the loaders will be the `in_datasets` that are available to the first processing plugin. It is important to note that the loader doesn't actually load any data, but loads the information required to access the data when a plugin is executed in a lazy fashion. The input data can be in variety of formats, with new loaders easily added to extend the range of input data that can be processed.

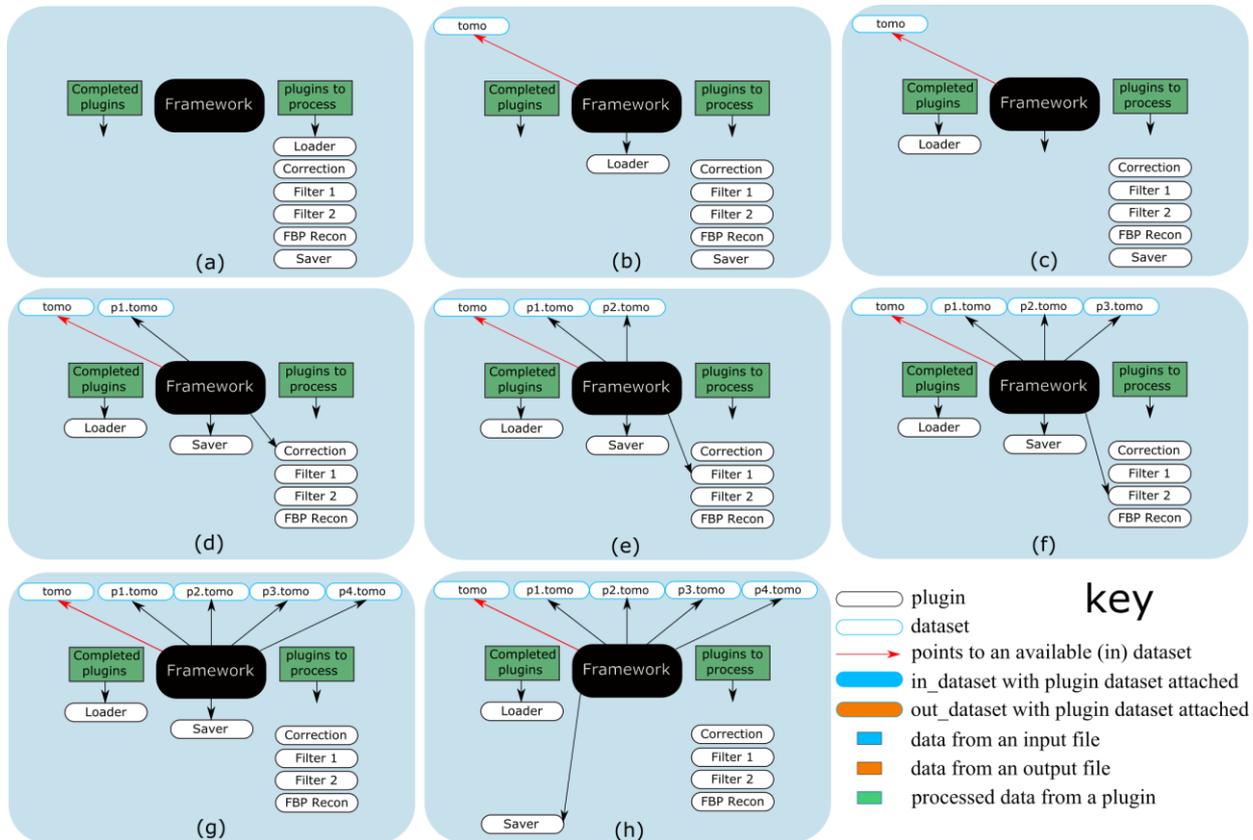

Fig. 5: Framework image 1 (setup phase). The framework is like a black box into which plugins are 'plugged-in', one after another in the order stated in a process list (a). The loader plugin is plugged into the framework (b), creating a dataset named 'tomo'. This `in_dataset` (distinguished by its red arrow) remains linked to the framework after the loader has completed (c). The saver plugin is then plugged-in, followed by the first processing plugin (d). The framework calls the setup method of the processing plugin, which provides the framework with the information it requires to create the associated `out_datasets`; in this case there is only one dataset and its name is also 'tomo', but since the framework retains the information that this `out_dataset` is linked to the first plugin, it has been distinguished with the name 'p1.tomo'. This step is repeated for all the processing plugins (e)-(g), with the final setup shown in (h).

The saver plugin deals with saving the data and, currently, only an HDF5 saver plugin is available. The HDF5 saver actually provides functionality for creating and saving HDF5 files. Although the saver plugin is the final plugin in the process list, it is actually called after the loader plugins have completed and retains a link with the framework until the processing chain has completed. Fig. 6 illustrates how the pipeline progresses through this setup phase, for a simple process list. This type of processing is typical in full-field, transmission tomography, with only one dataset being loaded and only one dataset name in the processing chain.

*3) Processing plugins*

Each processing plugin must contain:

- Number of `in_datasets` required
- Number of `out_datasets` required
- `in_dataset` and `out_dataset` parameter lists
- Setup method
- Processing method

Each processing plugin may optionally contain:

- Pre-process method
- Post-process method

Default methods for all mandatory methods, except the processing method, can be found in the BaseType class. It is possible then, that a plugin may only contain a processing method, if all the default methods are sufficient. The `in_datasets` and `in_datasets` lists contain the names of datasets that the plugin requires as input and output, and are set in the configurator when the process list is created and passed to the plugin at runtime. The number of each must equal the number specified by the plugin, and the input names must find a match in the available datasets list held by the framework. The framework will recognise these errors when it performs the plugin list check at the beginning of the processing chain, and will break the run before processing.

*4) Plugin datasets*

When a plugin is plugged-in, each dataset associated with that plugin will have a `plugin_dataset` attached to it for the duration of the plugin run. A `plugin_dataset` is an object containing extra information and functionality associated with the dataset for that particular plugin. A plugin dataset requires:

- Data access pattern
- Number of frames

The chosen data access pattern for a plugin must be available in the list of data access patterns associated with the dataset. If for example the pattern is set to 'SINOGRAM' and the number of frames is set to 8, then 8 sinograms will be passed to the main processing method at a time until all of the sinograms have been processed.

*5) Setup method*

The setup method is the first plugin method to be called when a plugin is plugged into the framework. At this stage, all the `in_datasets` and `out_datasets` required by the plugin will have been created, but the `out_datasets` will be empty, apart from a name, and will require populating with shape, axis labels and data access patterns. During the setup phase of the processing chain, after the setup method is called, the framework will attach a data file to the dataset, to complete the dataset, as illustrated in Fig 6. All the `plugin_datasets` that are associated with both `in_datasets` and `out_datasets` must also be populated with the required data access pattern and number of frames.

*G. Main Processing*

Once a plugin is executed, the methods are called in this order:

- Setup
- Pre-process
- Processing
- Post-process

The pre-process method is called once before the main processing step, the processing method is called in a loop until all data has been processed and the post-process method is called once (after an MPI barrier). Fig. 6 describes in detail how the framework handles the movement of data to and from a plugin during the main processing step, and Fig.7 illustrates the completion of the processing chain (See Fig. 10 on page 10 for a graphical description of a more complicated multi-modal processing chain).

## IV. PERFORMANCE ENCHANCEMENTS

Reading and writing data directly to and from file, rather than loading all data into RAM, provides a trade-off between data size and speed. The bottleneck in the MPI processing becomes the MPI I/O, as multiple cores compete to access the same file. In order to enhance the performance of Savu, optimisations were applied to the ordering of the data in memory (chunking), the MPI settings on the filesystem and the plugins themselves were optimised using an in-house MPI profiler developed from output logging information.

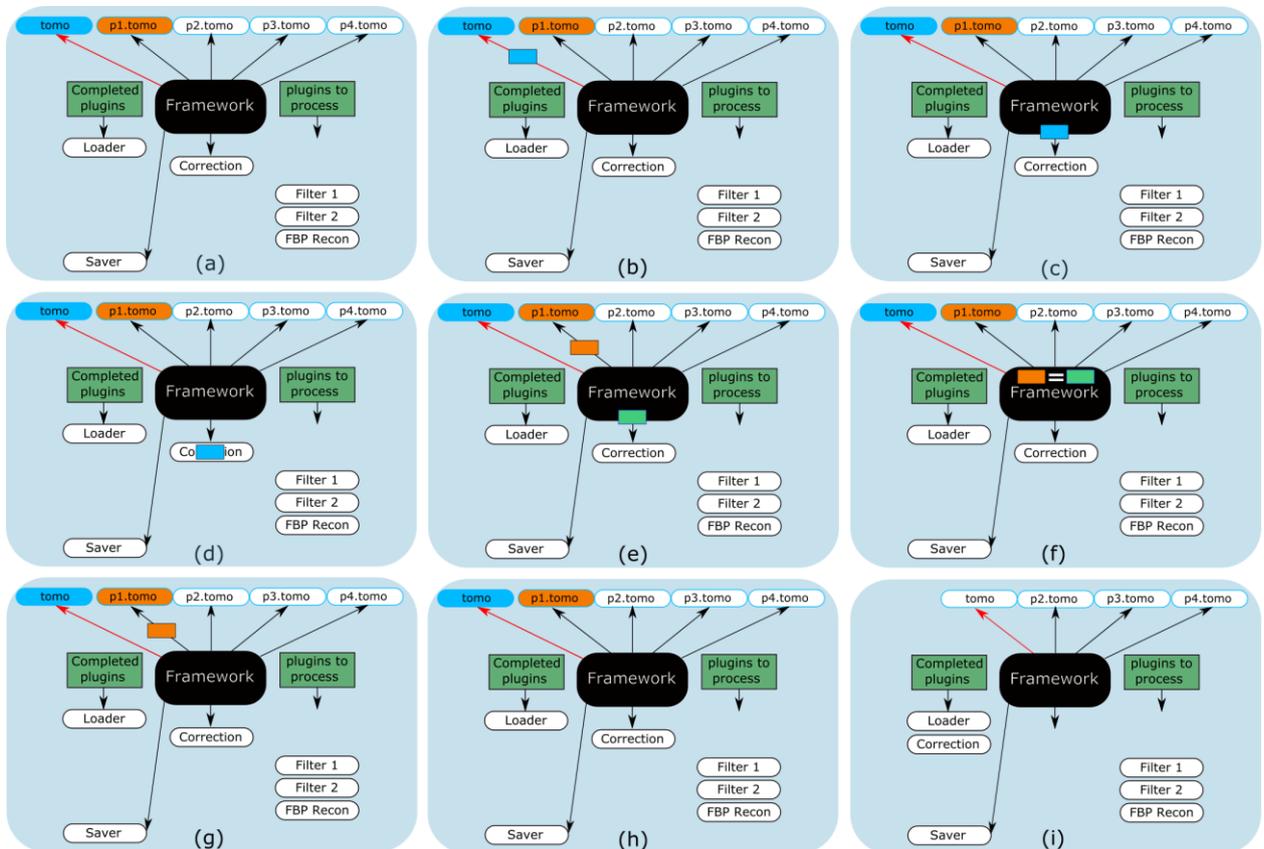

Fig. 6: Framework image 2 (main processing phase, key in Fig. 5). When a processing plugin is plugged-in to the framework, `plugin_datasets` are attached to the requested `in_dataset` and `out_dataset` (a). The framework then passes the requested *m* frames to the plugin (a)-(d), which then processes this data and returns the processed data back to the framework (e). The framework then grabs the aassociated (empty) data from the `out_dataset` and fills it with the processed data (f). The framework returns this data to the file (g),(h) and removes the `plugin_datasets` (i). The `in_dataset` is removed from the framework (after closing the associated file), and replaced by the `out_dataset` as they both have the same name.

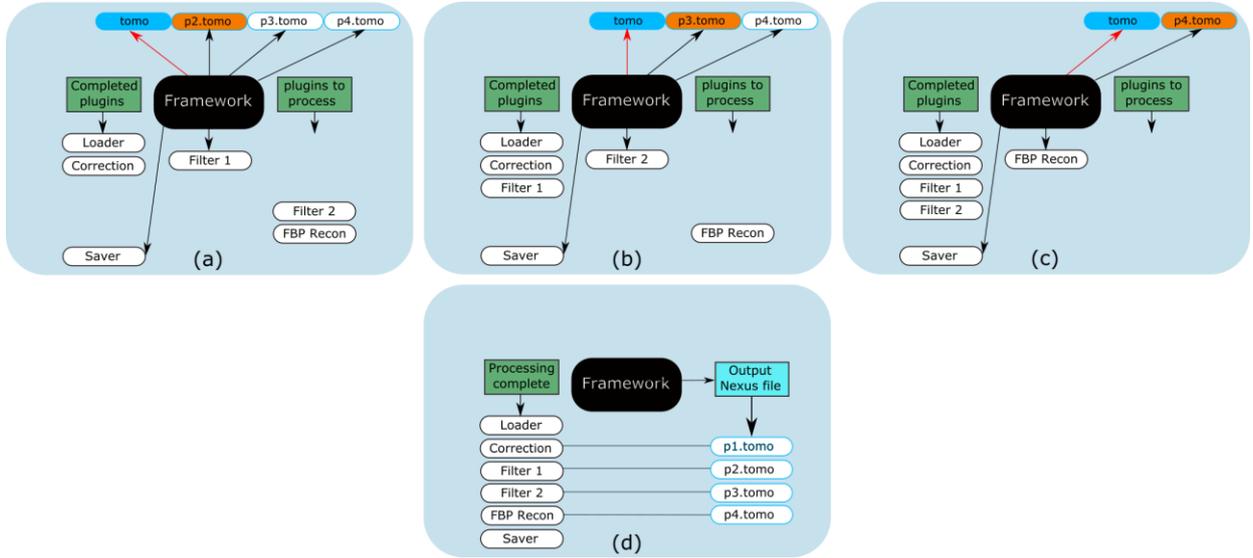

Fig. 7: Framework image 3 (main processing phase cont'd, key in Fig. 5). Each processing plugin is plugged-in one after the other (a)-(c), with the framework handling the organization of the data sets. Once the processing has completed, all the remaining data files are closed and an output NeXus file links them all together (d).

### A. Chunking

Chunking refers to the layout of the data in computer memory, and an optional 'chunk' value can be passed as a parameter when a HDF5 file is created. Savu uses the h5py python package [15] to interface the HDF5 software library, and passes the chunk value as a tuple with length $d$, where $d$ is the number of data dimensions. The dataset is then divided into equal 'chunks' of the specified size, with each chunk stored contiguously. The fewer chunks that need to be retrieved each time the data is accessed the better the performance will be.

HDF5 has a raw data chunk cache size, which determines how much data is read and written from file on each 'write' and 'read'. The default value is 1MB so, in this instance, the optimal performance would be achieved if 'chunk' was set to ensure that the 'chunk size' was 1MB. So, for example, if a 3-D dataset of single-precision floating point format (4 bytes per entry), had 'chunk' set to (1, 500, 500), then the chunk size would be 1 *x* 500 *x* 500 *x* 4 bytes = 1MB.

Each HDF5 dataset, created when the saver plugin is plugged-in to the framework, is linked to a Savu dataset, (see Fig...). Each Savu dataset can have more than one data access pattern associated with it, and this pattern may change during the processing chain. For instance, it may be created as an `out_dataset` in a plugin that processes in projection space, but used as an `in_dataset` in a plugin that processes in sinogram space. The framework does not know how a dataset will be processed until run-time, so optimal 'chunk' values are determined dynamically at runtime, in order to ensure that the data is organised in memory in the most efficient way.

Ensuring that minimum chunks are retrieved per access, whilst limiting chunks to a size of 1MB, is a delicate optimisation problem, which can have profound effects on performance. The formula that Savu uses for determining each chunking value is given below.

#### 1) The formula for optimised chunking

This optimization, is based on the first two data access patterns ('now' and 'next'), associated with the dataset, as it is rare that a dataset has more than two patterns associated with it. For a given pattern, a dimension is determined to be of type 'core', 'slice' or 'other' if it is a core dimension, a first slice dimension, or any other slice dimension respectively. Each dimension, $i$, is then allocated a start chunk value ($c_i$) based on the combination of these types for both 'now' and 'next' patterns. If the dimension is 'adjustable', then it is also assigned four other values: adjustment values for increasing ($\alpha_i^u$) and decreasing ($\alpha_i^d$), and upper ($\beta_i^u$) and lower bounds ($\beta_i^d$), see Table 1 for values.

Let $D$ denote the set of dimensions $(1, 2, \ldots, d)$, $D_c$ the set of core dimensions and $D_s$ the first slice dimension, then $D_a = D_c \cup D_s \subseteq D$ denotes the adjustable dimensions. Let $M$ denote the HDF5 chunk cache size (e.g. 1000000), then for $j \in D_a$, the final chunk value is calculated by,

$$c_j = \begin{cases} \alpha_j^u : & \prod_{\{i=1\}}^{d} c_i \leq M \\ \alpha_j^d : & \prod_{\{i=1\}}^{d} c_i > M \end{cases} \quad (1)$$

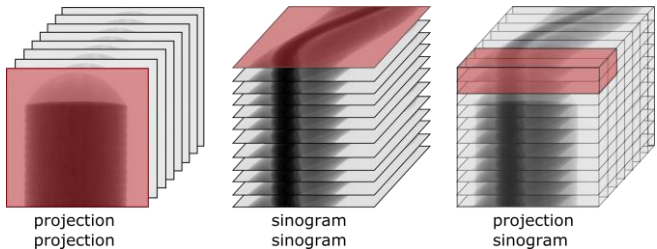

Fig. 8: An example of the most efficient way to chunk the data in memory (one chunk is red) for different now and next patterns. If one chunk was greater than the HDF5 chunk cache size (default 1MB) then it would be more efficient to reduce these chunks so they have size less than (as close to) or equal to 1MB.

| Now, Next | core | core | core | slice | slice | other |
|---|---|---|---|---|---|---|
| Next, Now | core | slice | other | slice | other | other |
| $c^0$ | $d$ | $\min(f,d)$ | 1 | $\min(f,d)$ | 1 | $1=c$ |
| $\beta^u$ | $d$ | $f_p$ | $d$ | $f_p$ | $d$ | |
| $\beta^d$ | 1 | 1 | 1 | 1 | 1 | |
| $\alpha^u$ | $c^0+a$ | $c^0+af$ | $c^0+a$ | $c^0+af$ | $c^0+a$ | |
| $\alpha^d$ | $c^0/2$ | $c_0-af$ | $c_0-a$ | $c_0-af$ | $c_0-a$ | |

Table 1: The initial values assigned to each dimension, based on their now and next patterns). Here $d$ is the number of dimensions, $f$ is the number of frames, and $f_p$ is the average number of frames processed by each MPI process. The values of $a$ and $b$ are given in the main text.

If the first condition in (1) is met, i.e. the chunking values are to be increased, then $D_a$ is ordered as $(D_c, D_s)$, otherwise $D_a$ is ordered as $(D_s, D_c)$. In calculating a and b, in the initial values, first consider that $a$ must satisfy the following equations,

$$c_j \leq b_j^u,$$
$$c_j - \prod_{\{k \in D \setminus \{j\}\}} c_k \leq M, \quad (2)$$

such that,

$$a = \left(\operatorname{argmin}_{\{a \in A\}}(\alpha_{j(a)}^u)\right), \quad (3)$$

where,

$$A = \{a : a \in N^0 \text{ and } \alpha_j^u(b) \leq \min(\beta_j^u, M - \prod_{\{k \in D-\{j\}\}} c_j)\} \quad (4)$$

and $b$ must satisfy the equations,

$$c_j \geq b_j^u,$$
$$c_j - \prod_{\{k \in D-\{j\}\}} c_k \leq M, \quad (5)$$

such that,

$$b = \operatorname{argmax}_{\{b \in B\}}(\alpha_j^u(b)), \quad (6)$$

where,

$$B = \{b : b \in N^0 \text{ and } \beta_j^l \leq \alpha_j^u(b) \leq M - \prod_{\{k \in D-\{j\}\}} c_j\} \quad (7)$$

*A. MPI settings*

After optimising the chunking parameters, the performance was still exceptionally slow on the GPFS filesystem and it was identified that only 1KB writes to file were being performed, rather than the expected 1MB. After investigating the MPI parameters available for tuning in the ROMIO implementation, the problem was solved by setting 'romio_ds_write' to 'disabled'.

*B. MPI profiling*

With the MPI I/O now performing well, performance focus was shifted to individual plugins. An MPI profiler is available with the Savu distribution, which allows visualisation of the entries in a log file after a Savu processing chain has completed. The output provides a simple visualisation of the time taken ++to perform each processing step, by each MPI process and is extremely useful in tuning individual plugins and process lists. An example output is given in Fig. 9.

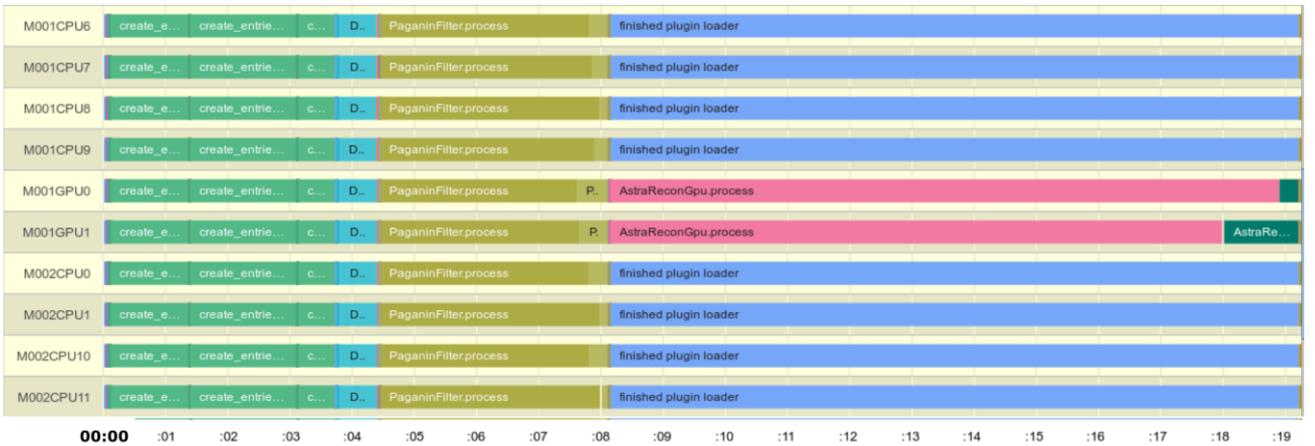

Fig. 9: A sample output from the Savu MPI profiler.

## V. CONCLUSION

The design choices and complexities required in developing a completely flexible system for processing tomography data have been presented. The initial concept of developing a scalable system, capable of running across multiple cluster compute nodes and processing terabytes of high dimensional data has been successfully designed and implemented. In addition, Savu is easy to install, develop and use; ensuring the latest developments in tomographic processing are available immediately and reducing the barriers for users to fully exploit available cluster compute resources.

The framework itself has been successfully deployed to the central cluster at Diamond Light Source, as well as the STFC central computing cluster SCARF. To date, Savu has been utilised on 5 different experimental stations at DLS, with several key improvements identified in comparison with older systems. On the full-field tomography branches I12 and I13 the flexibility that Savu enables, of processing in both projection and sinogram space as required, has led to the routine use of some phase contrast methods [16], which previously required manual intervention before reconstruction. Time resolved studies have also been transformed, by making use of Savu's multi-dimensional functionality to reconstruct full time series simultaneously, as opposed to reconstructing each time frame independently.

Finally, in mapping tomography, the integration of all required processing steps into one piece of software, and consequently the ability to run these processing steps as a single processing chain across large number of CPU cores, has led to a step change in the speed of processing from 6 hours, on the scientist's workstation, to 15 minutes across two nodes (40 cores) of the DLS cluster.

## VI. ACKNOWLEDGEMENT

The authors would like to thank Robert Atwood and Nghia Vo for being early adopters of Savu on the Full-Field imaging beamlines at Diamond Light Source. They would also like to thank Steven Price and Aaron Parsons for their work in converting their original code to use the Savu framework, and subsequently testing it with the Mapping beamlines at Diamond Light Source.

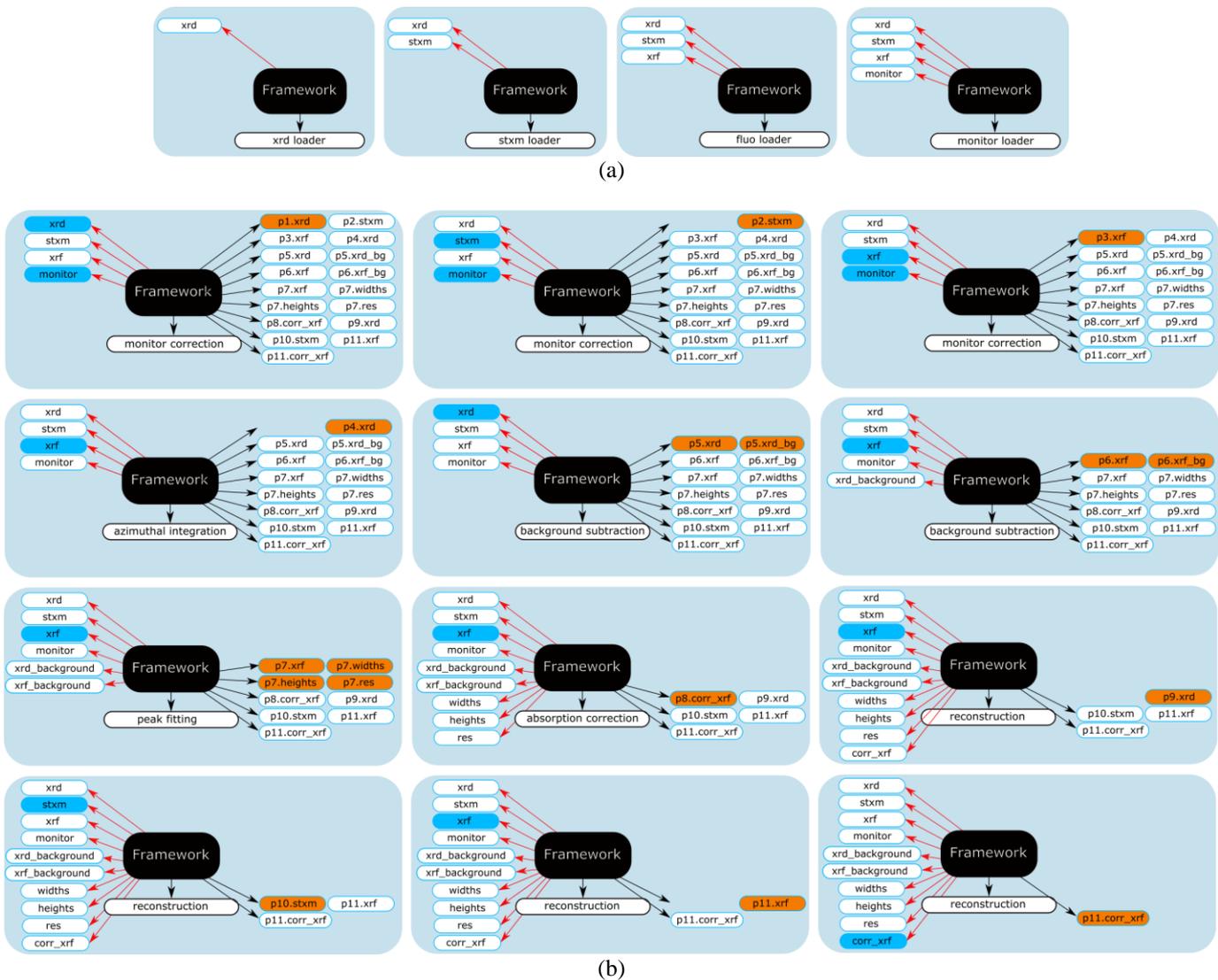

Fig. 10: Framework image 4 (Multi-modal data processing). Part (a) illustrates the setup phase of a processing chain requiring multiple loaders, resulting in multiple available `in_datasets` with unique names. Part (b) illustrates the main processing phase of a complicated, multi-modal processing chain and how the framework handles the datasets. The same plugin is applied multiple times to different datasets; some plugins require multiple `in_datasets` or `out_datasets` and new datasets (no previous dataset with the same name) are created as the processing chain is traversed.